\documentclass[journal,onecolumn]{IEEEtran}
\IEEEoverridecommandlockouts
\usepackage{cite}
\usepackage{amsmath,amssymb,amsfonts}
\usepackage{algorithmic}
\usepackage{pdflscape}
\usepackage{graphicx}
\usepackage{setspace}
\usepackage{subcaption}
\usepackage{tabu}
\usepackage[font=footnotesize,labelfont=bf]{caption}
\usepackage{textcomp}
\usepackage{xcolor}
\usepackage[inline]{enumitem}
\begin{document}
% \bstctlcite{IEEEexample:BSTcontrol}

\title{ \huge 
Testing Spintronics Implemented Monte Carlo Dropout-Based Bayesian Neural Networks
% Implemented on Spintronics-Based Architectures
% Computation-in-Memory
}

\author{\IEEEauthorblockN{Soyed Tuhin Ahmed\IEEEauthorrefmark{9}\IEEEauthorrefmark{2}, Kamal Danouchi\IEEEauthorrefmark{3}, Michael Hefenbrock\IEEEauthorrefmark{4}, Guillaume Prenat\IEEEauthorrefmark{3}, Lorena Anghel\IEEEauthorrefmark{3}, Mehdi B. Tahoori\IEEEauthorrefmark{2}\\}
\IEEEauthorblockA{\IEEEauthorrefmark{2}Karlsruhe Institute of Technology, Karlsruhe, Germany, \IEEEauthorrefmark{9}corresponding author, email: soyed.ahmed@kit.edu}
\IEEEauthorblockA{\IEEEauthorrefmark{3}Univ. Grenoble Alpes, CEA, CNRS, Grenoble INP, and IRIG-Spintec, Grenoble, France}
\IEEEauthorblockA{\IEEEauthorrefmark{4}RevoAI GmbH, Karlsruhe, Germany}
}
\maketitle

% \addtolength\abovedisplayskip{-0.6em}%
% \addtolength\belowdisplayskip{-0.6em}%
% \setlength{\textfloatsep}{0pt}

% add a starting sentence on importance and benefits of BayNN --> uncertainty estimation of predictions for functional safety

\begin{abstract}
% Bayesian Neural Networks (BayNNs) can inherently estimate uncertainty in prediction which allows informed decisions to be made.
Bayesian Neural Networks (BayNNs) can inherently estimate predictive uncertainty, facilitating informed decision-making.
Dropout-based BayNNs are increasingly implemented in spintronics-based computation-in-memory architectures for resource-constrained yet high-performance safety-critical applications. Although uncertainty estimation is important, the reliability of Dropout generation and BayNN computation is equally important for target applications but is overlooked in existing works. However, testing BayNNs is significantly more challenging compared to conventional NNs, due to their stochastic nature. In this paper, we present for the first time the model of the non-idealities of the spintronics-based Dropout module and analyze their impact on uncertainty estimates and accuracy. Furthermore, we propose a testing framework based on repeatability ranking for Dropout-based BayNN with up to $100\%$ fault coverage while using only $0.2\%$ of training data as test vectors.

% high endurance, high switching speed, and fast parallel computation of the BayNN operations.
\end{abstract}


\newcommand{\vx}{\ensuremath{\mathrm{\mathbf{x}}}}
\newcommand{\mW}{\ensuremath{\mathrm{\mathbf{W}}}}
%\newcommand{\vpsi}{\ensuremath{\mathrm{\bm{\psi}}}}

% \addtolength\abovedisplayskip{-0.6em}%
% \addtolength\belowdisplayskip{-0.6em}%
% \setlength{\textfloatsep}{0pt}
% \setlength{\belowcaptionskip}{-3pt}
% \setlength{\abovecaptionskip}{-1pt}
% \titlespacing\section{0pt}{12pt plus 4pt minus 2pt}{0pt plus 2pt minus 2pt}
% \titlespacing\subsection{0pt}{12pt plus 4pt minus 2pt}{0pt plus 2pt minus 2pt}
% \titlespacing\subsubsection{0pt}{12pt plus 4pt minus 2pt}{0pt plus 2pt minus 2pt}
% \titlespacing*{\subsection}{0pt}{1.0ex plus 1ex minus .2ex}{1.5ex plus .2ex}

\section{Introduction}
Bayesian Neural Networks (BayNNs) offer substantial benefits over conventional neural networks (NNs), particularly in safety-critical applications where reliability and confidence in prediction are paramount~\cite{tambon2022certify}. Unlike traditional NNs, BayNNs can inherently capture and estimate the \emph{uncertainty} of their predictions, enhancing decision-making under uncertain conditions. However, their implementation faces significant computational bottlenecks, especially on edge devices. Spintronics-based computation-in-memory (Spintronics-CIM) architectures are a promising solution for the hardware realization of BayNNs as they mitigate some of the inherent computational costs, balancing high-performance demands with the constraints of resource-limited devices.

Despite the advantages offered by Spintronics-CIM architectures, Spintronics memories are not without drawbacks due to their immature fabrication process and inherent non-deterministic behavior. 
Spintronics devices are susceptible to manufacturing defects such as pinholes in MgO layers, backhopping, side-wall redeposition, and magnetic coupling, causing instability of
magnetic states of ferromagnetic layers, and process variations that cause variability in magnetic properties~\cite{wu2020survey}. In addition, run-time variations and failures, including thermal variability lead to retention faults and read decision failures~\cite{wu2020survey}, significantly impacting BayNNs' accuracy and violating functional safety features,  
% One of the primary concerns is the presence of non-idealities that can adversely affect the accuracy of BayNNs~\cite{}, thereby violating functional safety features. These non-idealities in spintronics memories can reduce the accuracy and make the predictions overconfident.
% also affect the reliability of Dropout generation during Bayesian inference, 
a crucial aspect that has been overlooked in existing works~\cite{ahmed2023scale, ahmed2023spatial, ahemd_spindrop}. Addressing these reliability issues with proper periodic in-field testing is critical to ensuring the reliability of BayNNs in real-world safety-critical applications.

However, testing Dropout-based BayNN in Spintronics-CIM presents unique challenges, primarily due to the stochastic nature of its output. The same input can yield different output probability and uncertainty estimates, attributed to the random Dropout masks that create different sparse networks in each forward pass.
% ~\ref{fig:Dropout_prob}. 
This phenomenon is referred to as the \emph{repeatability problem}, 
% complicates the testing process, 
as it deviates from the deterministic output behavior seen in NNs.
Therefore, many of the existing testing approaches~\cite{meng2021self, Soyed_ITC, meng2022exploring, liu2020monitoring, chen2021line, luo2019functional, su2023testability, clue} cannot be used to test BayNNs as a constant deterministic set-point cannot be employed.

\begin{figure}
    \centering
    \includegraphics[width=0.55\linewidth]{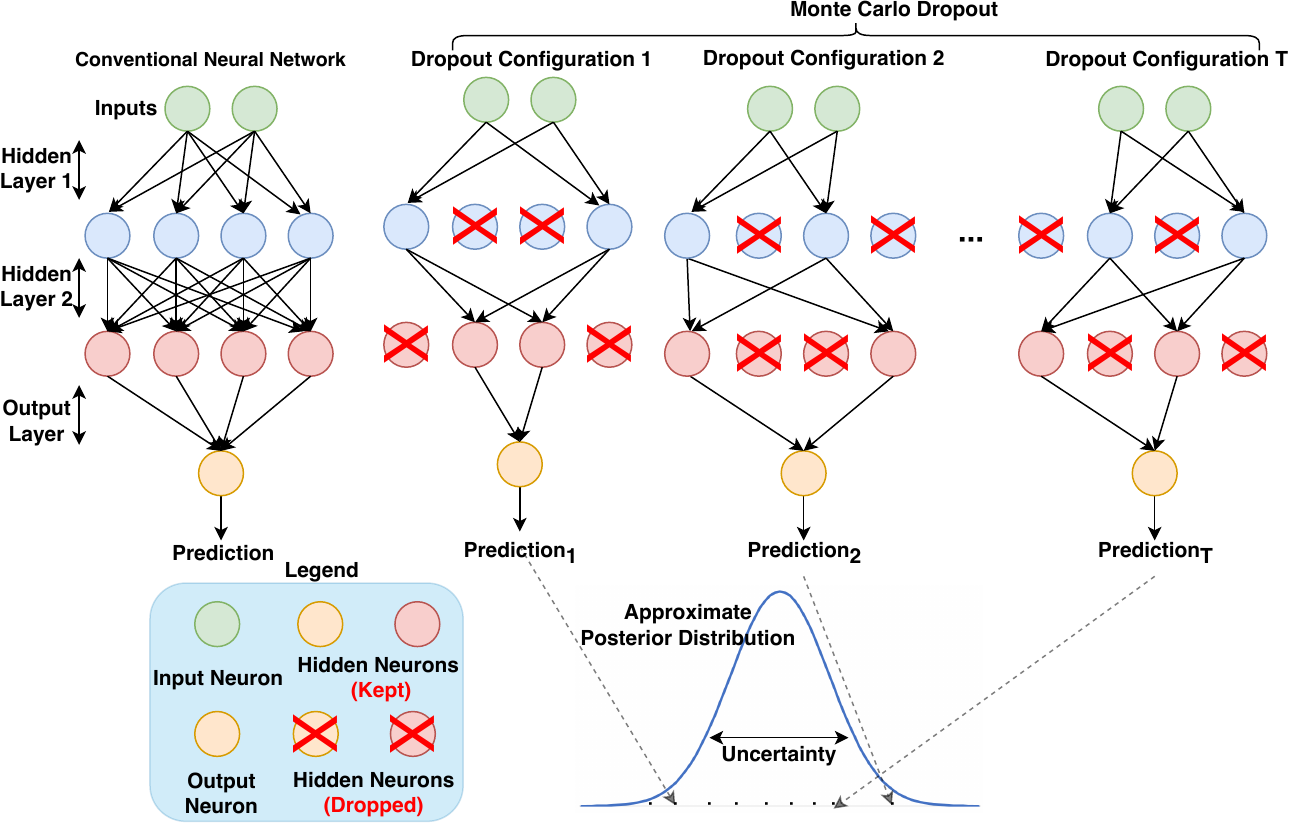}
    \caption{Conventional and Monte Carlo Dropout-based Bayesian inference, where the same input is applied to the network with T times with T different Dropout configurations to get the approximate posterior distribution.}
    \label{fig:MC_Dropout}
\end{figure}

Therefore, in this work, we propose a \emph{repeatability ranking-based} automatic test pattern generation (ATPG) and online testing framework to test BayNNs in Spintronics-CIM. The proposed framework is specifically designed to test defects, variations, and faults in Spintronics memories, buffer memory (that stores activations), and Spintronics-based Dropout modules. Furthermore, we model the non-idealities of dropout modules and perform fault injection analysis to evaluate their impact on the accuracy and quality of uncertainty estimates. We consider several design choices, such as time-multiplexing the dropout module within a layer and Dropout module sharing among different layers.
To the best of our knowledge, this is the first work that evaluates these aspects of BayNNs. 
% The proposed framework is specifically designed to self-test faults and variations in weights, activation functions, and the Dropout module of BayNNs. 
Furthermore, our approach treats the NN as a black box and is non-invasive. Thus, it is suitable to test pre-trained models from machine learning as a service (MLaaS) where NN models are treated as intellectual property (IP).

The remainder of this paper is organized as follows: Section~\ref{sec:Preliminary} gives a basic background related to our approach and discusses related works, in Section~\ref{sec:model} failure mechanisms and fault models of Spintronics-CIM are described, and in Section~\ref{sec:Proposed}
our proposed automatic test generation and testing framework is presented. Later, in Section~\ref{sec:Results} we evaluate our approach in detail, and in Section~\ref{sec:conclusion} the paper is concluded.

% The challenges of testing bayNN are test latency, coverage, FPR, MAC operation, and power consumption. We should cover all aspects. What is the baseline we can compare with?

% mention that self-testing test vectors need to be collected for each model post-training, but pre-deployment, they are stored in the hardware in secure off-chip memories like SRAM, and so on.

% The test is done online while NN is non-operational and periodically; thus, it is imperative that the individual test cost in terms of latency, power consumption, MAC operations, and memory consumption be minimal.

% We take the sum of uncertainty, as it allows uncertainty to be proportional to its uncertainty value. On the other hand, the sum-based evaluation will increase with the number of classes. By considering the uncertainty of all classes, our approach makes decisions based on the uncertainty of all classes rather than only the predicted classes. 

% The definition of good functional test vectors is that they are more sensitive to parameters or activation perturbations than others. A test vector that is more sensitive will detect more faults (high coverage), especially when the fault rate is low, i.e., the accuracy degradation is low.

% BNN is typically based on sampling-based techniques that involve the iterative generation of samples from posterior distributions.

\section{Preliminary}\label{sec:Preliminary}
\subsection{Dropout-Based Bayesian Neural Networks}
% Employing Dropout as an approximation for BayNNs has been shown to reduce the memory requirement to that of NNs.
% Nevertheless, 
% Deploying BayNNs on resource-constrained devices, edge applications, and in scenarios that demand high throughput is challenging.

% Unlike conventional NNs, BayNNs are probabilistic models. This means that BayNNs consider a probability distribution over their parameters instead of a single value point estimate, as done in NNs. The BayNN learning process involves the Bayesian framework, where the prior belief about the parameter is updated as more data is seen during training to get the posterior distribution for Bayesian inference. Unfortunately, the exact computation of the posterior distribution is intractable. Therefore, modern BayNNs resort to various approximation methods such as variational inference, the Laplace approximation, and the Monte Carlo (MC) dropout. Among them, MC-Dropout is attractive for Bayesian inference on resource-constrained devices because it has the same number of parameters as conventional NNs.

BayNNs, unlike traditional NNs, are probabilistic as they use parameter probability distributions rather than fixed point estimates. The learning incorporates Bayesian principles, updating prior beliefs with new data to estimate posterior distributions for inference. However, the exact computation of the posterior distribution is intractable, leading to reliance on approximations like variational inference, Laplace approximation, and Monte Carlo (MC) dropout. MC-Dropout is particularly suited for resource-limited devices due to its parameter count being equivalent to conventional NNs.

Bayesian inference with MC-Dropout involves passing the same input through the model $T$ times,
with the Dropout layer \textbf{\emph{turned on}} during inference. Therefore, each forward pass
requires sampling a different (independent) dropout mask with probability p, and the model has
a subset of active neurons, as shown in the Fig.~\ref{fig:MC_Dropout}. Consequently, each of the $T$ forward passes results in $T$ 
stochastic outputs, which is considered the posterior distribution. The final prediction
involves calculating the mean of the $T$ predictions in all dropout configurations. However, the variance represents the uncertainty of the predictions. In particular, a low variance value represents low uncertainty, and a high variance represents high uncertainty. In some cases, other methods are used to estimate uncertainty, but we use the variance-based method for a fair comparison.

\subsection{Spintronics Device and Bayesian Inference in Spintronics-CIM}

The core element of the Spintronics device is the Magnetic Tunnel Junction (MTJ). MTJ consists of two ferromagnetic layers – the Free Layer (FL) and the Reference Layer (RL) – separated by a thin insulating barrier. The resistance state of the device depends on the relative magnetization orientation of the ferromagnetic layers. Specifically, parallel (P) and anti-parallel magnetic orientation represent low resistance (LRS) and high resistance state (HRS), respectively.

In the Spintronics-CIM architecture, Spintronics cells are arranged in a crossbar array to enable MAC operations in an analog fashion with $O(1)$ time complexity~\cite{wang2020resistive}. The binary weights of the BayNN methods~\cite{ahemd_spindrop, ahmed2023scale, ahmed2023spatial} are encoded as the conductance state of the Spintronics with a one-time write operation before deployment.
During online operation, input vectors for a layer are converted to voltages using digital-to-analog converters (DACs) and are fed into the crossbars by activating multiple word-lines. The resulting output currents on the bit line of the crossbar represent the MAC results. The wordline current is sensed using analog-to-digital converters (ADCs). Afterward, operations such as bias addition, batch normalization, and binarization are performed on the digital outputs. The binary output becomes the input for the next layer and the operations are repeated. 

% MC-Dropout Implementation 
% Performing Dropout-based Bayesian inference on the CIM architecture requires performing different operations at the periphery depending on the method implemented. For example, the SpinDrop method requires
% randomly disabling each bit-line with a probability in each forward pass. The dropout module is implemented using the stochastic regime of the MTJs. Specifically, a series of SET and RESET is performed by applying a specific write current that represents the dropout probability.

Dropout-based Bayesian inference in the CIM architecture involves performing different operations in the periphery, depending on the implemented method. The SpinDrop method~\cite{ahemd_spindrop}, for instance, entails randomly disabling each bit-line of the crossbar (independently) with the probability $p$ in every forward pass. On the other hand, the SpatialSpinDrop~\cite{ahmed2023spatial} method, 
which is targeted for convolutional layers, randomly disables a group of bit-lines of the crossbar together with the probability $p$. However, the ScaleDrop method~\cite{ahmed2023scale} \emph{randomly scales} the current sum of the crossbar (after digitization) with probability $p$. 

The dropout module is implemented utilizing the inherent stochastic properties of spintronics devices, while the deterministic properties are used for BayNN weight storage. Specifically, CMOS transistors were integrated with the MTJ to allow control over the current and, hence, the switching probability of the MTJ. The Dropout procedure involved alternating "SET" and "RESET" operations to generate the Dropout bitstream. Following a "SET" write operation, the MTJ's state was read with a sense amplifier to validate the switching occurrence, thereby indicating the dropout signal. The MTJ was "RESET" to the P-state following the read operation in preparation for the next round of dropout signal production.

\subsection{Related Works}
In the literature, several works have proposed testing \emph{conventional} as well as \emph{spiking} NN~\cite{meng2021self, Soyed_ITC, meng2022exploring, liu2020monitoring, chen2021line, luo2019functional}. Although they are efficient, they do not account for multiple forward passes, or the stochasticity of Dropout-based BayNNs. Furthermore, some of the work, such as~\cite{meng2021self}, requires intensive fault injection studies that have a high computational cost that increases with the size of the model.
% In addition, several works focused on explicit testing of the convention NN~\cite{li2019rramedy, chen2021line, Soyed_ITC, luo2019functional} but their drawbacks are similar to self-test methods and 
Existing works often require the storage of a large number of test vectors, which can be as much as $10000$. Consequently, several test compression methods are proposed for spiking and conventional NN, such as~\cite{el2022compact, Soyed_ITC}. Furthermore, several works proposed to concurrently test conventional NN such as~\cite{gavarini2022open}. There are several other works for testing CIM architectures, as summarized in~\cite{su2023testability}. On the other hand, March-based algorithms~\cite{liu2016efficient} will require a large test time as BayNN implementations can have a significantly large number of memory cells.

% In work~\cite{clue}, the proposed cross-layer uncertainty estimator method (CLUE) considers both variations in conductance and algorithmic uncertainty. They proposed an ensemble of CIM process-based uncertainty estimations, where task processing is done in an analog processor, and uncertainty estimation is done in digital processors. They showed that with the process variation of CIM, uncertainty increases. However, their approach adds additional memory, latency, and energy consumption. Furthermore, they did not study the effect of faults on the digital processor that stores CLUE parameters and other types of faults that can affect NN predictions, such as faults in the weights and activation.

The cross-layer uncertainty estimator method (CLUE) in work~\cite{clue} proposed to estimate uncertainty due to conductance variations. They proposed an ensemble and knowledge distillation-based approach for uncertainty estimation for conductance variations.
% that is implemented in digital processors. However, tasks are processed on analog processors. 
Although this approach demonstrates increased uncertainty due to variations, it incurs additional memory, latency, and energy costs for each prediction. 
% Furthermore, the study does not explore the impact of faults on the digital processor storing CLUE parameters, or other fault types affecting NN predictions, such as faults in weight and activation.

In summary, Bayesian inference using Dropout-based approximation is different from conventional NN and spiking NN. Testing BayNN implemented in Spintronics-CIM presents unique challenges due to their stochastic output and multiple forward passes. Thus, existing work may not be applicable to test BayNN efficiently. Our work proposes a thorough evaluation of different non-idealities of Spintronics-CIM and an efficient testing method.

\section{Failure Mechanisms of BayNN in Spintronics-CIM}\label{sec:model}
\subsection{Failure Mechanisms of Spintronics Device}

The failure mechanisms for spintronics devices, as detailed in \cite{wu2020survey}, encompass:
\begin{enumerate*}
    \item \textbf{Manufacturing defects} which include front-end-of-line (FEOL) defects such as semiconductor impurities, crystal imperfections, and pinholes in gate oxides, and back-end-of-line (BEOL) defects like pinholes in MgO barriers, sidewall redepositions, and magnetic layer corrosion.
    \item \textbf{Extreme process variations} that lead to significant deviations in key MTJ parameters such as magnetic anisotropy, saturation magnetization, TMR ratio, and cross-sectional area, affecting both MTJs and transistors.
    \item \textbf{Magnetic coupling} phenomenon that pertains to stray magnetic fields from the ferromagnetic layers of neighboring MTJs or within an MTJ cell, affecting the stability of magnetization in the FL of MTJs.
    \item \textbf{STT-switching stochasticity} refers to the inherent stochastic nature of spintronics device switching, influenced by variability in incubation delay and actual switching time, leading to transient faults during write operations.
    \item \textbf{Thermal fluctuation} as a result of online temperature variations can significantly impact the magnetization reversal process in MTJs, causing retention and read decision faults. Due to these faults, the conductance state of a spintronics device can suddenly switch to another state.
\end{enumerate*}
The first two mechanisms generally result in permanent faults, while the last three are associated with transient faults in Spintronics devices. 

% The binary weights of the BayNN are encoded using STT-MRAM in crossbar arrays. 

\subsection{Failure Mechanisms of Buffer Memories}

In Spintronics-CIM, frequently updated results, such as intermediate activations, are stored in the register, flip-flops, or SRAM memories. In SRAM, permanent open and short faults can occur due to bridging defects that create an unwanted current that leads to a path between two nodes in the cell. Similarly, resistive open defects lead to an increase in the resistance of existing
paths within the cell~\cite{mirabella2021comparing}. Furthermore, transient faults can occur as a result of temperature and voltage fluctuations, as well as radiation particles striking memory cells.

% Moreover, for permanent faults, this is even more critical: a permanent fault in an activation would probably mean that the register or the primary output of a core executing the NN algorithm incurred a fault; since usually the NN cores are re-used multiple times during NN inference, this would not only impact a single activation but multiple activations.

% Actually, the complexity of the state-of-the-art devices makes exhaustive Fault Injection (FI) campaigns impractical and typically out of the computational capabilities.

% Finally,the IC goes through the extensive testing and post-silicon debug to check for bugs that escaped pre-silicon validation as well as manufacturing defects.

% STT-MRAM is a bi-stable device,

% Unlike other NVM technologies, STT-RAM does not exhibit dynamic write disturb faults~\cite{}. However, the main sources of soft errors in STT-RAM are read disturb, false read, and incomplete write faults. These faults are
% related to the uncertainty in the MTJ switching behavior. In addition to the faults
% discussed above, this section mainly focuses on incomplete writing faults.
\subsection{Failure Mechanisms of Spintronic-Based Dropout Module}
Due to the permanent and transient faults Spintronics device discussed before, the Dropout module can cause a word-line or a group of word-lines in the crossbar to be constantly active or inactive, or change the number of inactive word-lines. Furthermore, Dropout probability is highly sensitive to the switching current. Therefore, a small fluctuation in the switching current can cause variations in the Dropout probability.

\section{Proposed Approach}\label{sec:Proposed}

\begin{figure}
    \centering
    \includegraphics[width=\linewidth]{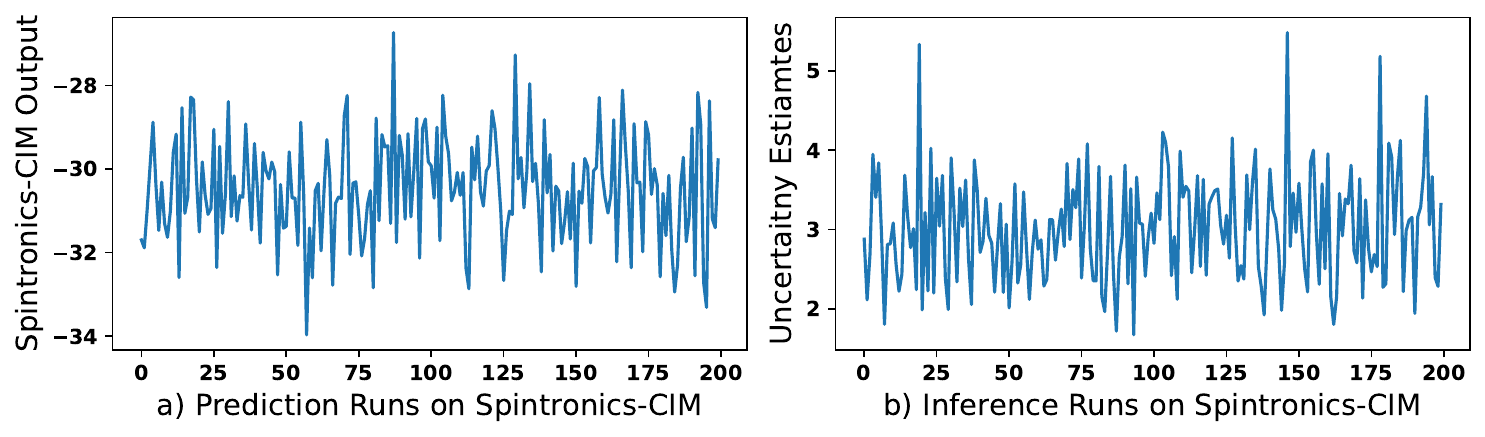}
    \caption{Stochasticity of a) Spintronics-CIM output (logits values) and b) uncertainty estimates for the same input for $200$ different predictions and inference runs, respectively.  }
    \label{fig:motivation}
    % \vspace{-1em}
\end{figure}

\subsection{Problem Statement}
Dropout-based BayNNs and their spintronics implementation represent a distribution of output over possible models, rather than a single model. Thus, the Spintronics-CIM outputs are different for the same input, as shown in Fig.~\ref{fig:motivation} (a). We refer to this characteristic as the repeatability or high-variance problem. Consequently, functionally testing BayNNs on Spintronics-CIM introduces several unique challenges due to the non-deterministic nature of their output due to the application Dropout, i.e., disabling the word-line of the Spintronics-based crossbar randomly.
% (irrespective of the stochastic behavior of Spintronics), 
% in line with the challenges of testing Spintronics devices.
% That means for the same input, BayNN predicts different predictions and uncertainty due to random dropout masks. 
Due to the repeatability problem, traditional functional testing-based approaches for testing Spintronics-CIM can lead to too many false positives or low true positive rates. This is because traditional approaches compare the expected output, label, or distribution of Spintronics-CIM to a pre-defined {deterministic} one without considering the stochasticity of BayNNs. Also, unlike the ensemble-based approach in~\cite{clue}, the uncertainty estimates of BayNN in Spintronics-CIM are also stochastic, even for the same input, as shown in Fig.~\ref{fig:motivation} (b). This creates an additional challenge in testing BayNNs in Spintronics-CIM based on their uncertainty.
% Therefore, estimation uncertainty due to faults and variations of Spintronics-CIM
% Thus, comparing the runtime status of BayNN in Spintronics-CIM to pre-defined output in a deterministic way is challenging, as it is not possible to predict the exact output for a given input.

Additionally, the input space of BayNNs can be very large, especially for models with many hidden layers. This is because Dropout-based BayNNs in Spintronics-CIM effectively create a separate sparse model for each forward pass, as shown in Fig.~\ref{fig:MC_Dropout}. Therefore, there are many possible inputs that the BayNN in Spintronics-CIM could be tested with. Exhaustively testing the model on all possible inputs would be computationally infeasible.
% expensive and would not be feasible for most applications.

Furthermore, the test cost of BayNN in terms of latency and power presents a significant challenge for resource-constrained devices or real-time applications. In these applications, the availability of the device is an important factor and cannot be unavailable for too long while a test operation is performed. However, Bayesian inference requires numerous forward passes to obtain a prediction for the input sample. Therefore, the number of test vectors should be minimal.
%\vspace{-\baselineskip}

\subsection{Automatic Test Generation Framework}

To address the challenges mentioned before, we propose a novel \emph{sample-based automatic test vector generation framework} to test BayNNs in Spintronics-CIM. In our approach, a small subset of training data is sampled based on their variance in the output. Here, variance is treated as a measure of the repeatability of the predictions. We hypothesized and observed that, despite the stochastic output of the model, the variance of inputs varies from one to another. Inputs with low variance are close to the deterministic model and are thus more suitable for testing BayNN in Spintronics-CIM, as they yield more interpretable outputs and uncertainties.

Our approach involves performing statistically significant repetitive inference runs for each training data sample. Subsequently, the variance in the uncertainty is calculated for each input. Additionally, training data is ranked based on their variance, with lower-variance samples receiving higher priority for selection. Lastly, several lower-variance training data points are stored in the hardware as test vectors.

In our approach, test vectors are sampled from the training dataset because we observed that the uncertainty of the distribution of training and validation data overlaps when no random data augmentations are applied.
% , as shown in Fig.~\ref{}. 
Therefore, it can be stated that the uncertainty of the prediction is the same regardless of the input data seen during the training, as long as they are of the same distribution. Therefore, no holdout data are required, which could otherwise reduce the available data for model training or validation. Consequently, our approach is particularly advantageous in scenarios with limited data.

\subsection{Proposed Fault Detection Approach}

For fault detection, we hypothesize that as the spintronics device deviates from its initial state or is faulty, the uncertainty of the prediction increases. Our hypothesis is grounded in the fact that as the distribution of input to the Spintronics-CIM shifts away from the training distribution, e.g., due to random noise introduced by the sensor or by dynamic environmental conditions such as rain, snow, or fog, the uncertainty of prediction increases
% (see Fig.~\ref{fig:uncer_obser}).
% A similar observation is 
as demonstrated in several existing works~\cite{gawlikowski2023survey, ahmed2023scale, ahmed2023spatial, ahemd_spindrop}. In scenarios where faults and variations occur on the Spintronics cells storing BayNN weights and buffer memories storing activations of hidden layers, inputs to subsequent layers will also change from their expected values, effectively creating \emph{intermediate out-of-distribution} data, even though initial input to the Spintronics-CIM remains within the in-distribution range. Therefore, the uncertainty of the prediction is expected to increase.

Our objective is to detect these changes in prediction uncertainty as a means of testing the BayNN implemented in Spintronics-CIM. We define a predefined range for the uncertainty of the BayNN given the test vectors. The BayNN model is classified as faulty if the uncertainty distribution changes from the predefined distribution. Otherwise, it is not faulty.

We fit the uncertainty distribution of the test vectors to a Gaussian distribution, as they are mathematically well-defined and easy to work with. Afterward, we evaluate the mean $\mu$ and standard deviation $\sigma$ of the uncertainty distribution to estimate the baseline (fault-free) range of the prediction uncertainty. Specifically, we define the two bounds, $b_1$ and $b_1$, representing the upper and lower bounds of the distribution based on the empirical rule of probability. The rule states that $99.7\%$ of the values of a Gaussian distribution are within three standard deviations of the mean. Therefore,  $b_1$ and $b_1$ are defined as $\mu + 3\times\sigma$ and $\mu - 3\times\sigma$, respectively.

Lastly, during each online testing phase of the Spintronics-CIM, the test vectors are applied sequentially to the model as \emph{test queries}. If the uncertainty of the prediction is above or below the boundaries, the model is classified as faulty. Therefore, in our approach, only lightweight checks are required in the model output.
Afterward, thorough testing is required, for example, to localize the faults and perform re-training or re-calibration to mitigate the impact of faults and variations.

% Furthermore, in our approach, the test vectors and boundaries are stored in the hardware. During online testing, the test vectors are applied sequentially to the model as \emph{test queries}.

\subsection{Reduction of False Positives Rate}

A theoretically sound approach to address the repeatability problem and reduce the false-positive rate would be to test BayNNs based on expected output. Specifically, predictions and uncertainties are derived from the mean of multiple inference results, with each inference result obtained after multiple forward passes, as described earlier. However, this approach is impractical. For example, if the expected uncertainty is determined after $10$ inferences, each requiring 20 forward passes, the total computation for a single test vector reaches $200$ forward passes, which is prohibitively expensive.

% To further reduce the false-positive rate, , 
Therefore, we propose a low-cost \emph{vote-based approach}. Specifically, unlike in work~\cite{Soyed_ITC}, multiple test queries contribute to the fault identification process. However, to minimize test costs in terms of latency and power, we limit the length of the query sequence to the minimum when a close-to-ideal (less than 10\%) false positive rate is achieved.

\section{Evaluation}\label{sec:Results}
\subsection{Simulation Setup}\label{sec:sim_setup}
The proposed method is evaluated across three state-of-the-art Spintronics-CIM-implemented dropout-based BayNN methods: SpinDrop~\cite{ahemd_spindrop}, SpatialSpinDrop~\cite{ahmed2023spatial}, and ScaleDrop~\cite{ahmed2023scale}. All methods are implemented with the widely used ResNet-18 topology and the CIFAR-10 benchmark dataset. The scalability of the method is evaluated on a more difficult biomedical semantic segmentation task and a U-Net topology.

The test vector is generated after performing 200 repetitions of inference with each inference 20 Monte Carlo samples of the Dropout mask were done. For fault detection, a positive test query length of four was used and $100$ test vectors were used. We have performed the Monte Carlo fault simulation with a $1000$ fault injection for each fault or variation rate. In total, we have performed $60,000$ fault injection campaigns at different locations of Spintronics-CIM as mentioned in Section~\ref{sec:model}.

The proposed method evaluates fault coverage, which states how many injected faults are detected. Faults are treated as benign if the accuracy does not degrade noticeably. Otherwise, they are critical faults. Fault coverage for critical faults represents true positive rates (TPR) and without faults represents false positive rates (TPR).

% MC runs, MC-fault simulations, number of FI injection campaigns, Model used, evaluation methods like FPR, TPR, Fault coverages

\subsection{Fault Models For Spintronics-CIM-based BayNN}

Permanent faults in spintronics cells and buffer memory are modeled as stuck-at faults, as done typically, with logic values always fixed at '0' or '1'~\cite{su2023testability}. In binary BayNNs, these translate to weights and activations being persistently stuck at '-1' or '1'. In the Dropout module, the permanent fault results in a constant Dropout mask of '0' or '1'. Thus, a word-line in the crossbar can always be inactive or active. Similarly, the widely used bit-flip fault model is used for transient faults. In the bit-flip fault model, the logic value in the spintronics cells and the buffer memory randomly shift from '0' to '1' or vice versa. For binary BayNNs, this means random flipping of logic values between '-1' and '1'. In the Dropout module, this manifests itself as random bit flips in the Dropout mask. Thus, a  word-line that was originally inactive can be active, and vice versa.

% Permanent faults are commonly modeled as stuck-at faults (SAF) where the logic value of an STT-MRAM cell or buffer memory is always '0' or '1'~\cite{}. In the context of binary BayNNs, it translates to the logic values of the STT-MRAM cell that stores weights and buffer memory that stores activations always stuck at '-1' or '1'. Also, in the context of the Dropout module, the permanent fault is modeled as the logic value Dropout mask always stuck at either '0' or '1'. 

% Similarly, transient faults are commonly modeled as bit-flip faults, where the logic value of an STT-MRAM cell or buffer memory randomly flips from '0' to '1' or the other way around. In binary BayNNs, this means that the logic values of the STT-MRAM cells and buffer memory randomly flip from '-1' to '1' and vice versa.
% Also, in the context of the Dropout module, the transient fault is modeled as the random bit flipping of the logic values Dropout mask.

% Furthermore, existing works~\cite{} show that the conductance variations of STT-MRAM in a crossbar array cause variations in the current sum (MAC results). Also, work in~\cite{} modeled the conductance variation as an additive and multiplicative Gaussian noise. Thus, conductance variations of STT-MRAM cells are modeled as additive and multiplicative Gaussian noise at the MAC results.

Furthermore, existing works show that conductance variations can be modeled as additive and multiplicative Gaussian variations~\cite{kim2020efficient} and
variations in the resistive states of MRAM devices lead to variations in the current sum. 
Thus, based on these works~\cite{ahmed2023design}, conductance variations in spintronics devices are modeled as additive and multiplicative Gaussian variations in the MAC result.

Lastly, due to fluctuations in the switching current of the Dropout module, the Dropout probability can vary. Based on works~\cite{ahemd_spindrop}, variation in the Dropout probability is represented with a Gaussian distribution, the mean representing the original Dropout probability.

\subsection{Fault Sensitivity Analysis of BayNN on Spintronics-CIM}

\begin{figure}
    \centering
    \footnotesize
    \includegraphics[width=0.95\linewidth]{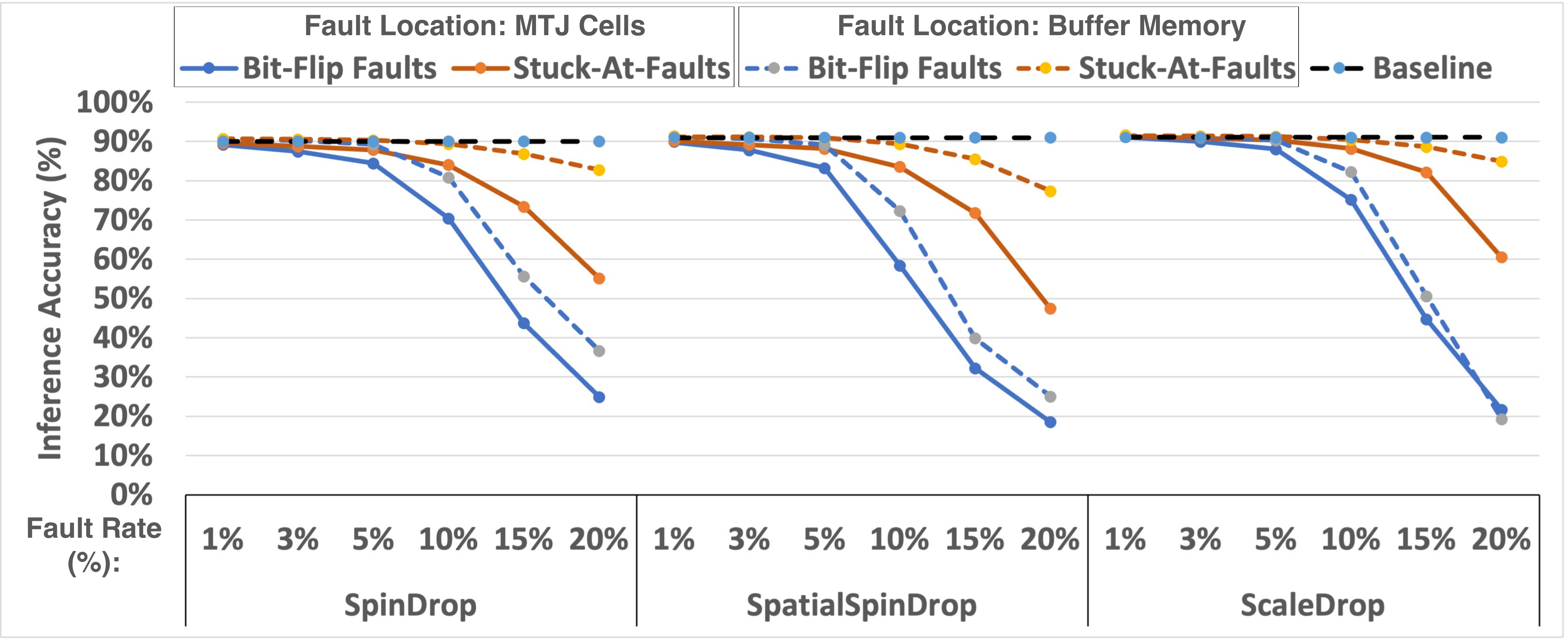}
\caption{
Impact of Inference accuracy of Spintronics BayNNs with different bit-flip and stuck-at-fault rates, compared to a baseline without faults.
% Impact of Inference accuracy of various Spintronics implemented BayNN
% SpinDrop, SpatialSpinDrop, and ScaleDrop 
% with various bit-flip and stuck-at-fault rates. The baseline (without faults) is shown for comparison.
}
    \label{fig:acc}
\end{figure}

\begin{figure}
    \centering
    \includegraphics[width=0.9\linewidth]{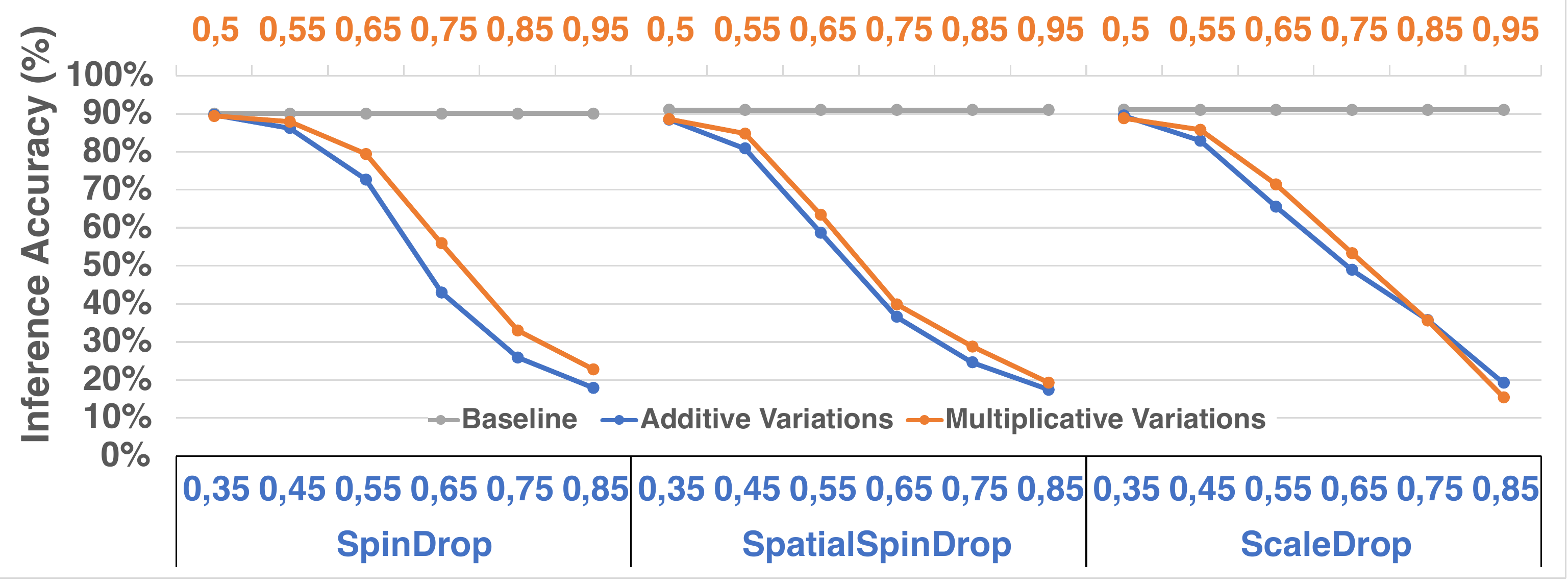}
    \caption{Comparison of the impact of inference accuracy of Spintronics BayNNs with conductance variations relative to a fault-free baseline.}
    \label{fig:acc_var}
    % \vspace{1em}
\end{figure}

% BayNNs are inherently more fault-tolerant compared to NN. 
BayNNs implemented in Spintronics-CIM can withstand up to $5\%$ of stuck-at and bit-flip faults in MTJs and buffer memories, as shown in Fig.~\ref{fig:acc}. In particular, BayNNs are more fault-tolerant in the case of stuck-at faults compared to bit-flip faults. Specifically, they can withstand up to $10\%$ of stuck-at-faults in buffer memories that store binary activations.
In terms of BayNN methods, scale-dropout-based BayNNs are more fault-tolerant and can tolerate up to $15\%$ of stuck-at-faults in buffer memories. In contrast, the inference accuracy gradually decreases with the additive and multiplicative types of MTJ conductance variations, as shown in Fig.~\ref{fig:acc_var}. 
% Therefore, at some fault rate, the BayNN accuracy remains stable and is considered \emph{ benign}.
Nevertheless, all BayNN methods show an overall trend of a reduction in accuracy as the fault rate increases. Therefore, this emphasizes the necessity for fault detection to ensure the functional safety and reliability of BayNNs.

\subsection{Analysis of Fault Coverage}

\begin{figure*}
    \centering
    \includegraphics[width=\linewidth]{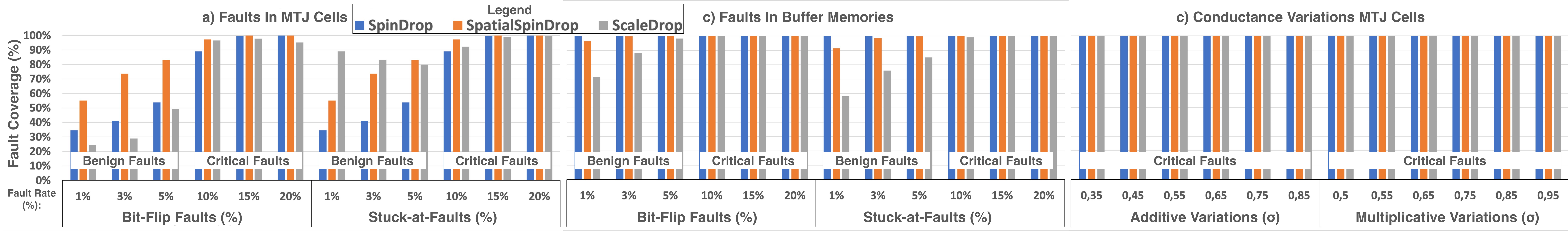}
     \caption{Fault coverage of proposed approach on various Spintronics implemented BayNN methods under varying bit-flip and stuck-at faults rate a) affecting Spintronics cells that store weights, b) buffer memories that store intermediate activation, and c) different conductance variations in Spintronics.}
     \label{fig:cov}
     %\vspace{-2em}
\end{figure*}

As shown in Figs.\ref{fig:cov} (a) and (b), our proposed approach can predominantly achieve $100\%$ fault coverage for critical faults on MTJs and buffer memories of Spintronics-CIM. Similarly, our proposed approach can consistently achieve $100\%$ fault coverage for conductance variations of spintronics devices, as depicted in Fig.\ref{fig:cov} (c). In the worst case, our approach can achieve a fault coverage of $89.10\%$. Consequently, our approach is more effective in detecting faults in the buffer memories and conductance variations. Specifically, even when the accuracy drop is marginal, i.e., $0.30\%$, our proposed approach can achieve $100\%$ fault coverage. In terms of BayNN methods, our approach is particularly effective with the SpinDrop and SpatialSpinDrop methods, which are more fault-sensitive compared to the ScaleDrop method.
% Especially for the SpinDrop and SpatialSpinDrop methods, as they are more fault-sensitive compared to the ScaleDrop method. 
Note that the fault coverages mentioned here are TPR and our approach can consistently achieve an ideal $100\%$ TPR value. 
% Also, when faults occur in Spintronics cells in sensitive locations on the crossbar, a small number of faults, for example, $\sim 0.1\%$, can cause a large accuracy degradation. In this case, our approach can also detect them with $100\%$ coverage. However, a thorough analysis of these aspects is beyond the scope of the paper and can be explored in future work.
% for critical faults represents the true positive rate, and 

Furthermore, our approach can also achieve up to $100\%$ fault coverage for benign faults. Detecting benign faults before they catastrophically degrade accuracy is beneficial, especially for faults, such as retention faults, that accumulate over time.
% Even for benign faults, our approach can consistently achieve $100\%$ fault coverage.

% \subsection{Analysis of Fault Coverage Due to Variations}
% We found that the ResNet-18 BayNN model is robust at lower conductance variation rates and that these rates are considered benign.
% Similarly, our proposed approach can consistently achieve $100\%$ fault coverage for all BayNN methods implemented in Spintronics-CIM, as depicted in Fig.\ref{fig:cov} (c). 
% Specifically, even when the accuracy drop is marginal, i.e., $0.30\%$, our proposed approach is highly effective in detecting MTJ conductance variations. 

% \begin{figure}
%     \centering
%     \includegraphics[width=\linewidth]{figures/FI_Weights.pdf}
%     \caption{Fault coverage of proposed approach on various Spintronics implemented BayNN methods, i.e., SpinDrop, SpatialSpinDrop, and ScaleDrop,  under varying bit-flip and stuck-at faults rate affecting MTJ cells.}
%     \label{fig:fi_weights}
% \end{figure}

% \begin{figure}
%     \centering
%     \includegraphics[width=0.8\linewidth]{figures/FI_activations.pdf}
%     \caption{Fault coverage of various Spintronics implemented BaNNs while faults occurred in buffer memories that store intermediate activation.}
%     \label{fig:fi_acts}
% \end{figure}

% \begin{figure}
%     \centering
%     \includegraphics[width=\linewidth]{figures/FI_variations.pdf}
%     \caption{Fault coverages in Spintronic-implemented BaNNs with conductance variations in STT-MRAM.}
%     \label{fig:fi_var}
% \end{figure}

\begin{figure}
    \centering
    \footnotesize
    \includegraphics[width=0.90\linewidth]{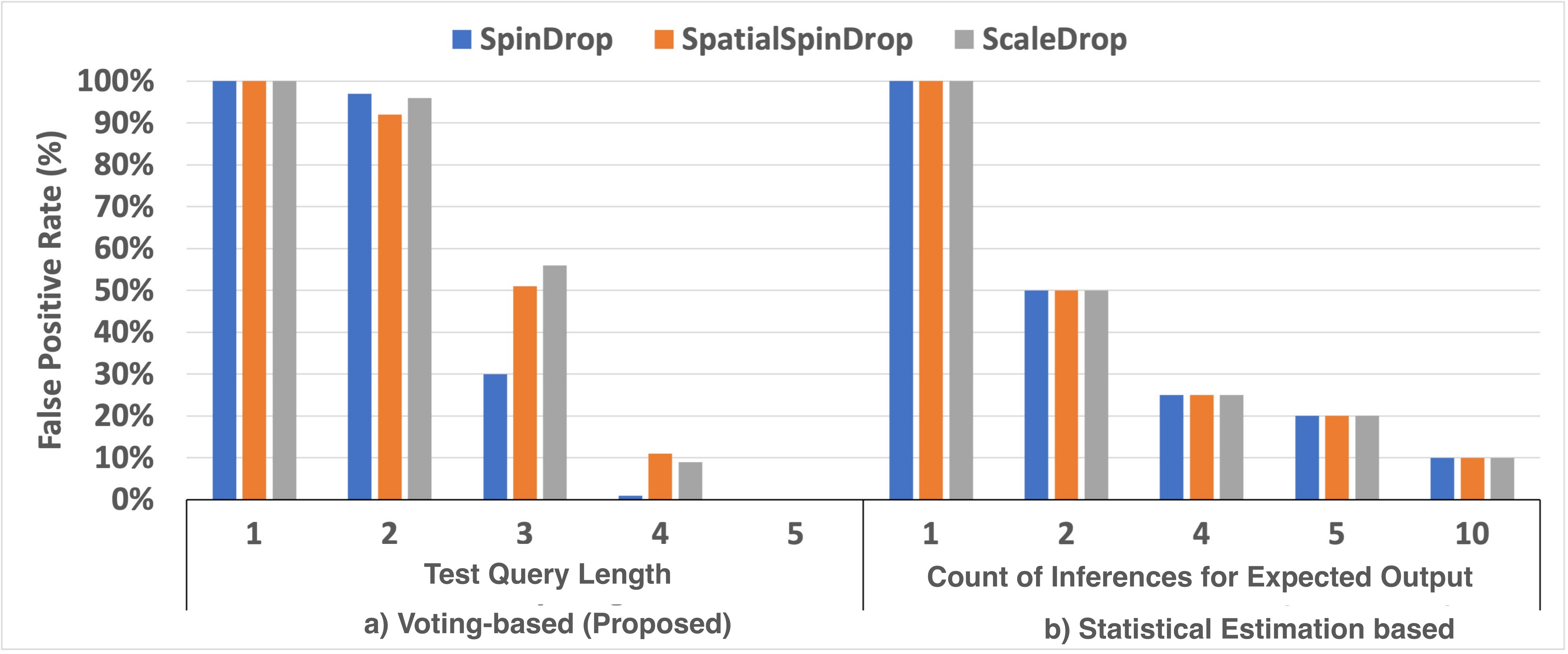}
    \caption{False positive rate (lower the better) of a) proposed voting-based approach, and b) theoretically grounded estimation-based approach.}
    \label{fig:FPR}
    % \vspace{-1em}
    %\vspace{-\baselineskip}
\end{figure}

\begin{figure}
    \centering
    \footnotesize
    \includegraphics[width=0.75\linewidth]{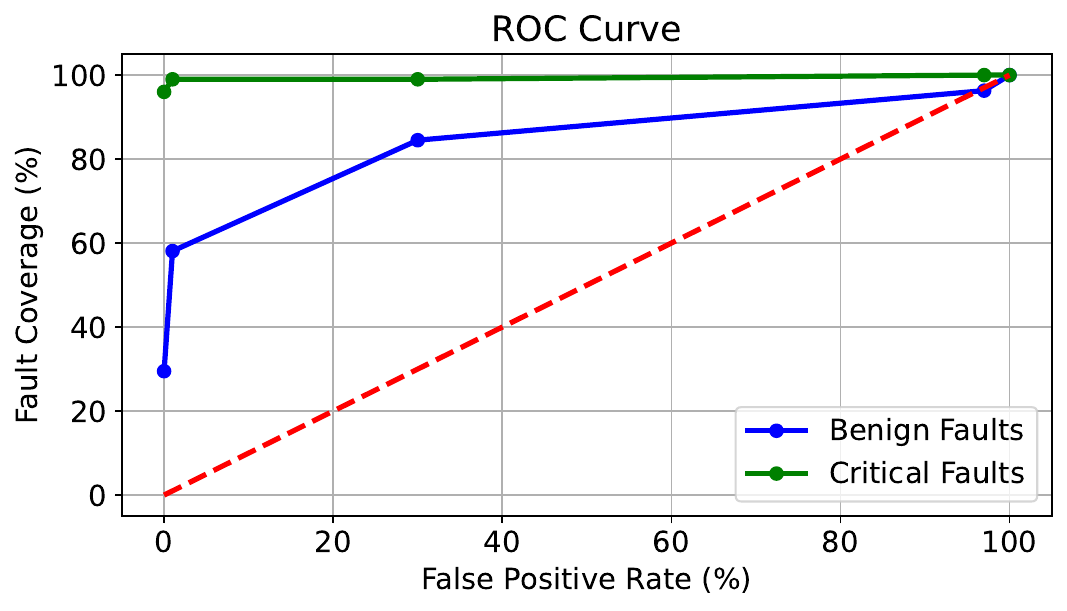}
    \caption{ROC curves for benign and critical faults with varying positive test query lengths. 
    % The diagonal line is for reference which represents random chance. 
    % , including a reference diagonal line which represents random chance.
    % The ROC curves of the proposed approach for benign and critical faults with different positive test query lengths. The red diagonal line is for reference which represents random chance. 
    }
    \label{fig:roc}
\end{figure}

\subsection{Analysis of False Negative Rate (FPR)}
Although achieving high fault coverage is important in the case of critical faults and conductance variations, achieving a low FPR is equally important in reducing false alarms. As shown in Fig.~\ref{fig:FPR}, our approach gradually reduces the FPR to an ideal value of $0\%$ with a positive test query length of $5$. As mentioned in Sec.~\ref{sec:sim_setup}, evaluation of test coverage performed with a positive test query length of $4$, which results in an acceptable FPR of $10\%$ and less. 

There is a trade-off between fault coverage and FPR given by the positive test query length, as shown in the Receiver Operating Characteristic (ROC) curve in Fig.~\ref{fig:roc} for the SpinDrop method. In an ROC curve, the closer a curve is to the top-left corner, the better the performance. With our proposed approach, both the curves for benign and critical faults perform well because they are above the random diagonal line. Specifically, the curve for critical faults does not change with FPR. This suggests that our approach can achieve $100\%$ fault coverage even with a positive test query length of $1$ and is better at detecting critical faults than benign ones. In contrast, the coverage of the benign fault is highly dependent on the positive test query length. In practice, we suggest using a positive test query length of $4$ as it can detect a sufficiently large number of benign faults and the majority of critical faults at low FPR.

\subsection{Scalability to Larger Dataset and Model}
% cifar100 and semantic segmentation application
We showed that our approach is highly effective in detecting faults and variations in Spintronics-CIM
% MTJs, buffer memory, and conductance variation 
for a large ResNet topology (18 layers) with different Dropout-based BayNNs. However, scalability to an even larger model and a harder dataset is important. Specifically, in the U-Net model for the biomedical semantic segmentation task, our approach can achieve a similar fault coverage of $100\%$ for bit-flip and stuck-at-faults in MTJ cells. Similar performance is also observed for faults in other locations of Spintronics-CIM.
% \vspace{-\baselineskip}
% buffer memories and conductance variation.

% \subsection{}

\subsection{Analysis of Non-ideal Dropout Module}
We found that BayNNs~\cite{ahemd_spindrop,ahmed2023scale,ahmed2023spatial} are robust to faults (stuck-at and bit-flip) and variation in the switching current of the Dropout modules and do not impact the inference accuracy with up to a $20\%$ fault rate. Similarly, uncertainty estimates of BayNNs are generally not affected. However, in the SpinDrop and ScaleDrop methods, bit-flip and stuck-at-0 faults lower the uncertainty, that is, making the BayNN \emph{overconfident}. In these scenarios, the faults are considered critical, and our approach can achieve $100\%$ coverage. Furthermore, if the Dropout modules are shared within a layer and all the other layers, then a fault in the Dropout module affects all the neurons of a layer and other layers. In this case, stuck-at faults make all the word-lines of Spintronics-CIM always inactive or active and make the uncertainty estimates \emph{zero}, i.e., the same as conventional NNs. In this case, our approach can also achieve $100\%$ coverage.

% \paragraph{Bit-flips of Dropout Mask}

% \paragraph{Stuck-at faults on Dropout Mask}

% \paragraph{Variation on Dropout Probability}

% \subsection{Empirical Evidence for the Hypothesis}
% show results-based random sampled images, and so on.
%\vspace{-1.em}

\subsection{Overhead Analysis and Comparison to Related Works}

In terms of memory overhead, our approach requires $0.31$ MB of memory to store test vectors. Furthermore, for a single test query, our approach consumes energy of $2$, $0.68$, and $0.18$ uJ for the SpinDrop, SpatialSpinDrop, and ScaleDrop methods, respectively. Here, energy consumption depends on the specific implementation of BayNN. For a BayNN status check, the overall energy consumption required depends on the number of test queries. In the worst case, our method requires $200$, $68$, and $18$ uJ energy, respectively, for $100$ test queries. In comparison, the statistical estimation-based approach requires $10\times$ energy consumption for a test query to achieve $10\%$ FPR.

As mentioned earlier, to the best of our knowledge, this paper proposes the first work on testing BayNN on Spintronics-CIM. The other related work~\cite{clue}, although not directly comparable, estimates the uncertainty of the model due to process variations using an ensemble-based approach. They do not report fault coverage, but an $80.4\%$ reduction in calibration error (a measure of uncertainty) is reported. Also, their work has a memory overhead of $0.45$ MB and an energy consumption of $6.02$ uJ per test query. This results in $602$ uJ for the $100$ test vectors in an explicit testing scenario.
% Therefore, our method requires $x\%$ less memory and $y\%$ less power consumption.

\section{Conclusion}\label{sec:conclusion}

In this paper, we propose for the first time a test generation and online testing framework for Dropout-based BayNN implemented in Spintronics-CIM. Also, fault analysis of different non-idealities of Spintronics-CIM is presented. Our approach can consistently detect $100\%$ of the critical faults at different locations. Furthermore, our approach requires only $0.2\%$ training data as test vectors and a simple check at the BayNN output. 

\bibliographystyle{IEEEtran}
\bibliography{bibliography}

\begin{thebibliography}{10}
\providecommand{\url}[1]{#1}
\csname url@samestyle\endcsname
\providecommand{\newblock}{\relax}
\providecommand{\bibinfo}[2]{#2}
\providecommand{\BIBentrySTDinterwordspacing}{\spaceskip=0pt\relax}
\providecommand{\BIBentryALTinterwordstretchfactor}{4}
\providecommand{\BIBentryALTinterwordspacing}{\spaceskip=\fontdimen2\font plus
\BIBentryALTinterwordstretchfactor\fontdimen3\font minus \fontdimen4\font\relax}
\providecommand{\BIBforeignlanguage}[2]{{%
\expandafter\ifx\csname l@#1\endcsname\relax
\typeout{** WARNING: IEEEtran.bst: No hyphenation pattern has been}%
\typeout{** loaded for the language `#1'. Using the pattern for}%
\typeout{** the default language instead.}%
\else
\language=\csname l@#1\endcsname
\fi
#2}}
\providecommand{\BIBdecl}{\relax}
\BIBdecl

\bibitem{tambon2022certify}
F.~Tambon, G.~Laberge, L.~An, A.~Nikanjam, P.~S.~N. Mindom, Y.~Pequignot, F.~Khomh, G.~Antoniol, E.~Merlo, and F.~Laviolette, ``How to certify machine learning based safety-critical systems? a systematic literature review,'' \emph{Automated Software Engineering}, vol.~29, no.~2, p.~38, 2022.

\bibitem{wu2020survey}
L.~Wu, M.~Taouil, S.~Rao, E.~J. Marinissen, and S.~Hamdioui, ``Survey on stt-mram testing: Failure mechanisms, fault models, and tests,'' \emph{arXiv preprint arXiv:2001.05463}, 2020.

\bibitem{ahmed2023scale}
S.~T. Ahmed, K.~Danouchi, M.~Hefenbrock, G.~Prenat, L.~Anghel, and M.~B. Tahoori, ``Scale-dropout: Estimating uncertainty in deep neural networks using stochastic scale,'' \emph{arXiv pp arXiv:2311.15816}, 2023.

\bibitem{ahmed2023spatial}
------, ``Spatial-spindrop: Spatial dropout-based binary bayesian neural network with spintronics implementation,'' \emph{arXiv preprint arXiv:2306.10185}, 2023.

\bibitem{ahemd_spindrop}
S.~T. Ahmed, K.~Danouchi, C.~Münch, G.~Prenat, L.~Anghel, and M.~B. Tahoori, ``Spindrop: Dropout-based bayesian binary neural networks with spintronic implementation,'' \emph{IEEE JETCAS}, vol.~13, no.~1, pp. 150--164, 2023.

\bibitem{meng2021self}
F.~Meng, F.~S. Hosseini, and C.~Yang, ``A self-test framework for detecting fault-induced accuracy drop in neural network accelerators,'' in \emph{26th ASP DAC}, 2021.

\bibitem{Soyed_ITC}
S.~T. Ahmed \emph{et~al.}, ``Compact functional test generation for memristive deep learning implementations using approximate gradient ranking,'' in \emph{2022 IEEE International Test Conference (ITC)}, 2022.

\bibitem{meng2022exploring}
F.~Meng and C.~Yang, ``Exploring image selection for self-testing in neural network accelerators,'' in \emph{ISVLSI}.\hskip 1em plus 0.5em minus 0.4em\relax IEEE, 2022.

\bibitem{liu2020monitoring}
Q.~Liu, T.~Liu, Z.~Liu, W.~Wen, and C.~Yang, ``Monitoring the health of emerging neural network accelerators with cost-effective concurrent test,'' in \emph{DAC}.\hskip 1em plus 0.5em minus 0.4em\relax IEEE, 2020.

\bibitem{chen2021line}
C.-Y. Chen \emph{et~al.}, ``On-line functional testing of memristor-mapped deep neural networks using backdoored checksums,'' in \emph{2021 IEEE ITC}, 2021.

\bibitem{luo2019functional}
B.~Luo, Y.~Li, L.~Wei, and Q.~Xu, ``On functional test generation for deep neural network ips,'' in \emph{DATE}.\hskip 1em plus 0.5em minus 0.4em\relax IEEE, 2019.

\bibitem{su2023testability}
F.~Su \emph{et~al.}, ``Testability and dependability of ai hardware: Survey, trends, challenges, and perspectives,'' \emph{IEEE Design \& Test}, 2023.

\bibitem{clue}
M.~Lee, A.~Lu, M.~Mukherjee, S.~Yu, and S.~Mukhopadhyay, ``Clue: Cross-layer uncertainty estimator for reliable neural perception using processing-in-memory accelerators,'' in \emph{IJCNN}, 2023.

\bibitem{wang2020resistive}
Z.~Wang, H.~Wu, G.~W. Burr, C.~S. Hwang, K.~L. Wang, Q.~Xia, and J.~J. Yang, ``Resistive switching materials for information processing,'' \emph{Nature Reviews Materials}, 2020.

\bibitem{el2022compact}
S.~A. El-Sayed, T.~Spyrou, L.~A. Camu{\~n}as-Mesa, and H.-G. Stratigopoulos, ``Compact functional testing for neuromorphic computing circuits,'' \emph{IEEE TCAD}, 2022.

\bibitem{gavarini2022open}
G.~Gavarini, D.~Stucchi, A.~Ruospo, G.~Boracchi, and E.~Sanchez, ``Open-set recognition: an inexpensive strategy to increase dnn reliability,'' in \emph{28th IOLTS}.\hskip 1em plus 0.5em minus 0.4em\relax IEEE, 2022, pp. 1--7.

\bibitem{liu2016efficient}
P.~Liu, Z.~You, J.~Kuang, Z.~Hu, H.~Duan, and W.~Wang, ``Efficient march test algorithm for 1t1r cross-bar with complete fault coverage,'' \emph{Electronics Letters}, vol.~52, no.~18, pp. 1520--1522, 2016.

\bibitem{mirabella2021comparing}
N.~Mirabella, M.~Grosso, G.~Franchino, S.~Rinaudo, I.~Deretzis, A.~La~Magna, and M.~S. Reorda, ``Comparing different solutions for testing resistive defects in low-power srams,'' in \emph{LATS}.\hskip 1em plus 0.5em minus 0.4em\relax IEEE, 2021.

\bibitem{gawlikowski2023survey}
J.~Gawlikowski, C.~R.~N. Tassi, M.~Ali, J.~Lee, M.~Humt, J.~Feng, A.~Kruspe, R.~Triebel, P.~Jung, R.~Roscher \emph{et~al.}, ``A survey of uncertainty in deep neural networks,'' \emph{Artificial Intelligence Review}, vol.~56, no. Suppl 1, pp. 1513--1589, 2023.

\bibitem{kim2020efficient}
H.~Kim, J.-H. Bae, S.~Lim, S.-T. Lee, Y.-T. Seo, D.~Kwon, B.-G. Park, and J.-H. Lee, ``Efficient precise weight tuning protocol considering variation of the synaptic devices and target accuracy,'' \emph{Neurocomputing}, vol. 378, pp. 189--196, 2020.

\bibitem{ahmed2023design}
S.~T. Ahmed, M.~Mayahinia, M.~Hefenbrock, C.~M{\"u}nch, and M.~B. Tahoori, ``Design-time reference current generation for robust spintronic-based neuromorphic architecture,'' \emph{ACM JETC}, vol.~20, no.~1, 2023.

\end{thebibliography}

\end{document}